\documentclass[doublecol]{epl2}
\usepackage{soul}
\usepackage{graphicx}
\usepackage{amsmath}
\usepackage{amssymb}
\usepackage{amsfonts}
\usepackage{subcaption}
\usepackage{bm}
\usepackage{xcolor}
\usepackage{cleveref}

\newcommand{\ra}{\operatorname{Ra}}
\newcommand{\pr}{\operatorname{Pr}}

\newcommand{\ste}{\operatorname{St}}
\newcommand{\Thfl}{T_{\text {\tiny HFL}}}
\newcommand{\Th}{T_{\text {\tiny H}}}
\newcommand{\Tc}{T_{\text {\tiny C}}}

\title{Heat Transfer in Phase Change Materials with Multiple Fin Insertion}
\shorttitle{Heat-Transfer \& Phase Change Materials with multi-fin configurations} 
\author{Paolo Proia\inst{1} \and Mauro Sbragaglia\inst{2} \and Giacomo Falcucci\inst{1,3}}
\shortauthor{P. Proia \etal}
\institute{
     \inst{1} Department of Enterprise Engineering “Mario Lucertini”, Tor Vergata University of Rome, Via del Politecnico 1 00133
    Rome Italy \\\\
    \inst{2} Department of Physics and INFN, Tor Vergata University of Rome, Via della Ricerca Scientifica, 1 00133 Rome, Italy\\\\
    \inst{3} Department of Physics Harvard University 33 Oxford Street 02138 Cambridge Massachusetts USA\\\\
 }
\abstract{We leverage 3D numerical simulations to study phase change materials (PCMs) cells under the effect of buoyancy forces. The solid PCM is heated from a source boundary, triggering melting.
The source features multiple solid fins that protrude into the PCM cell; the impact of the fins and their number is investigated by designing and testing equivalent (in terms of heating power) finless and single fin simulations. For each configuration, the performance is quantified via the total molten substance in time. The designs were also tested for different values of the non-dimensional numbers encoding relevant properties.
We confirm that fins increase the melting performance and find that single fin configurations are sub-optimal since a layout with multiple fins takes advantage of interstitial spaces, melting the substance more efficiently. The results also indicate that fins should be properly spaced, as closeness can result in overlapping, thus interfering, molten areas.
}
\begin{document}
\maketitle
\section{Introduction}
\begin{figure*}[t!]
    \centering
    \captionsetup[subfigure]{singlelinecheck=off,justification=raggedright}
    \begin{subfigure}[t]{0.5\linewidth}
        \caption{}
        \centering
        \begin{minipage}{0.45\linewidth}
            $$t^\star=0.00$$
            \includegraphics[width=\linewidth]{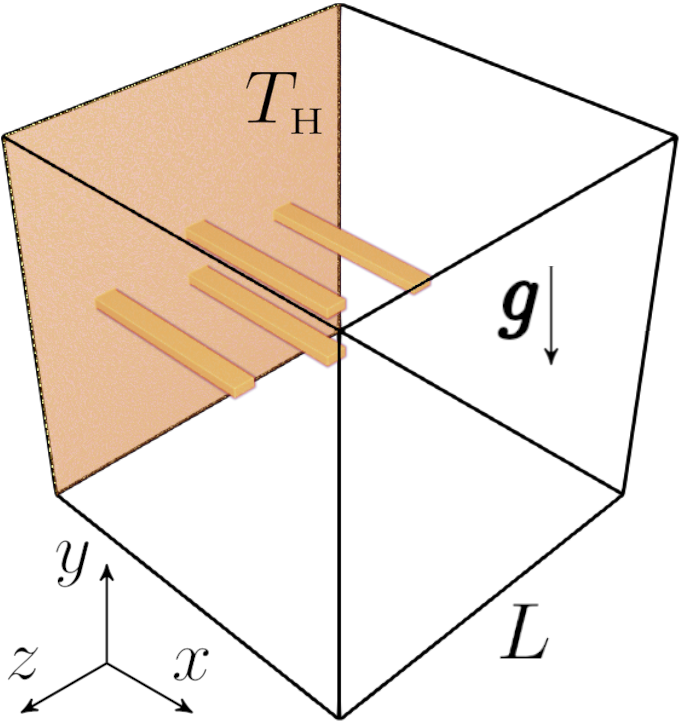}
        \end{minipage}
        \hfill
        \begin{minipage}{0.45\linewidth}
            $$t^\star=0.05$$
            \includegraphics[width=\linewidth]{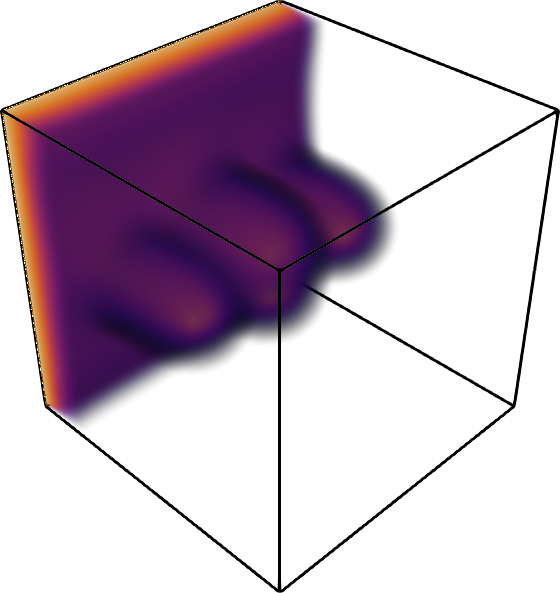}
        \end{minipage}\\
        \begin{minipage}{0.45\linewidth}
            $$t^\star=0.15$$
            \includegraphics[width=\linewidth]{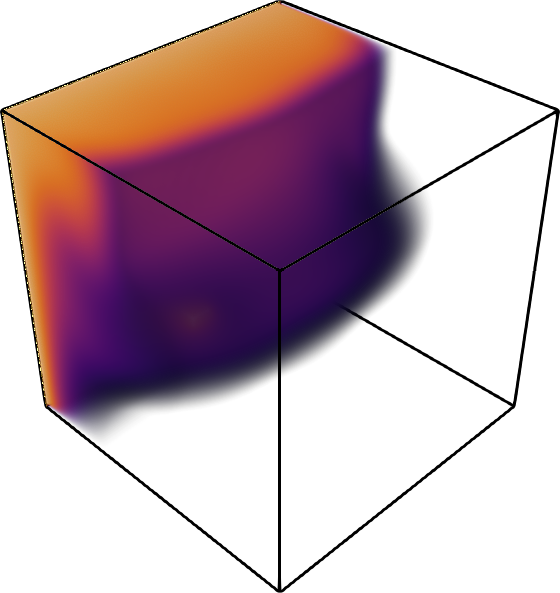}
        \end{minipage}
        \hfill
        \begin{minipage}{0.45\linewidth}
            $$t^\star=0.25$$
            \includegraphics[width=\linewidth]{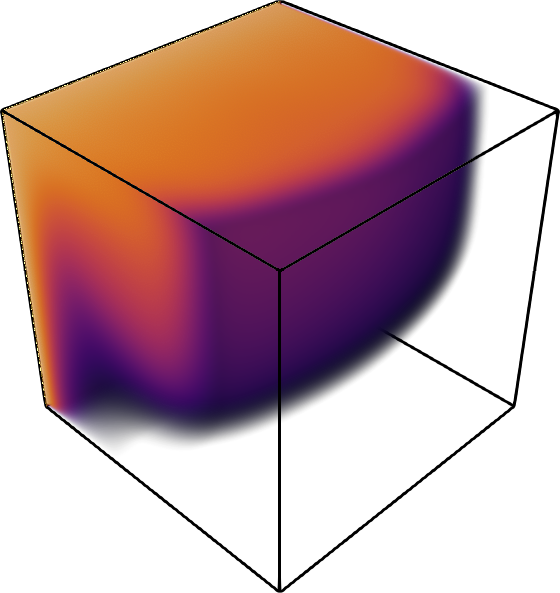}
        \end{minipage}\\
        \vspace{1em}
        \begin{minipage}{\linewidth}
            \centering
            \includegraphics[width=\linewidth]{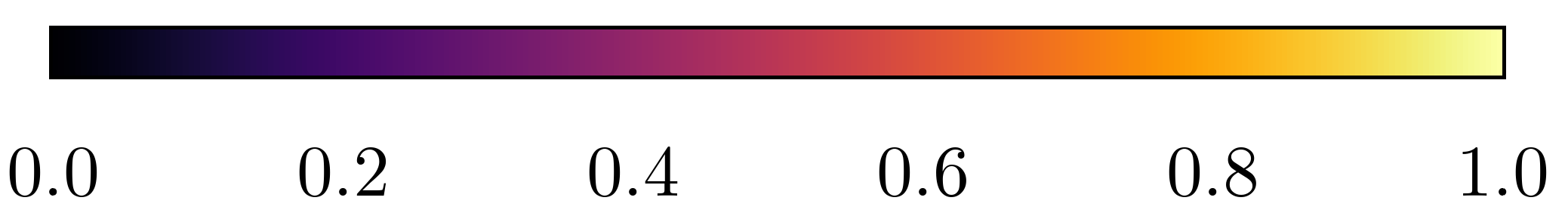}\\
            \vspace{-0.5em}$T$
        \end{minipage}
        \begin{minipage}{\linewidth}
            \vspace{1em}
            \centering
            \begin{minipage}[t]{0.45\linewidth}
                \raggedright (c)\\\vspace{0.5em}
                \includegraphics[width=\linewidth]{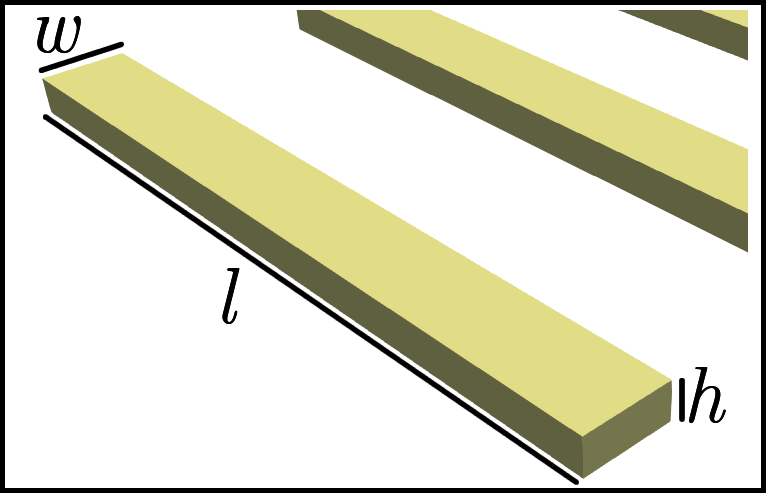}
            \end{minipage}
            \hfill
            \begin{minipage}[t]{0.45\linewidth}
                \raggedright (d)\\\vspace{0.5em}
                \includegraphics[width=\linewidth]{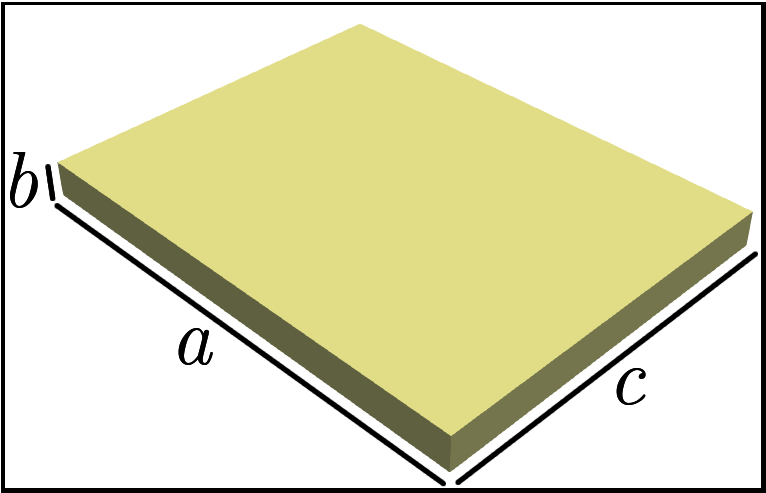}
            \end{minipage}
        \end{minipage}
    \end{subfigure}
    \hfill
    \begin{subfigure}[t]{0.45\linewidth}
        \caption{}
        \centering
        \begin{minipage}{0.4\linewidth}
            \centering
            \includegraphics[width=\linewidth]{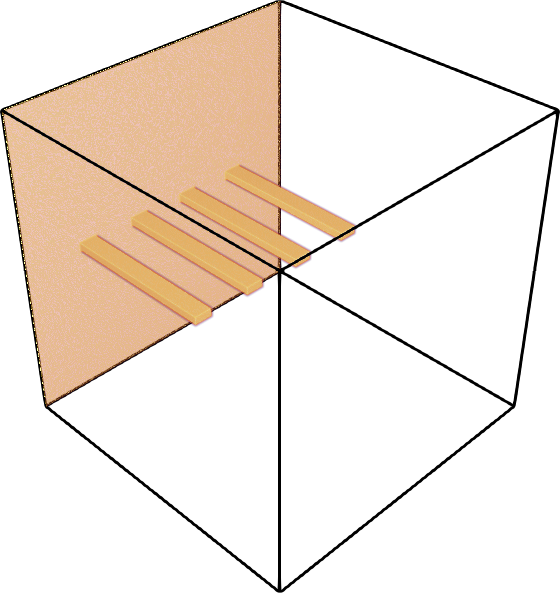}
            i) line
        \end{minipage}
        \hspace{0.1\linewidth}
        \begin{minipage}{0.4\linewidth}
            \centering
            \includegraphics[width=\linewidth]{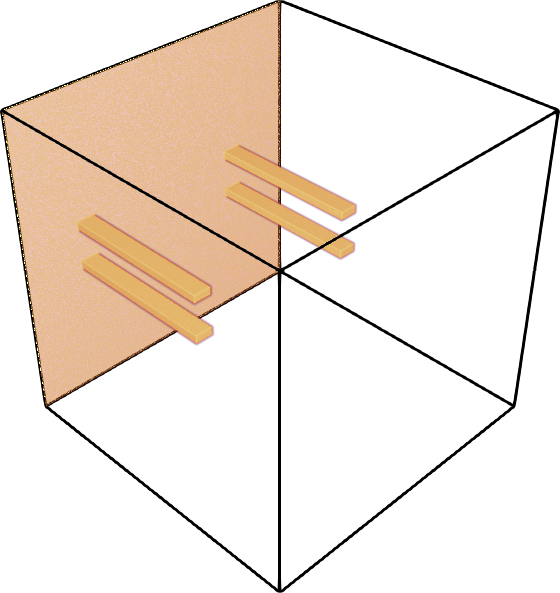}
            ii) rect
        \end{minipage}\\
        \vspace{0.5em}
        \begin{minipage}{0.4\linewidth}
            \centering
            \includegraphics[width=\linewidth]{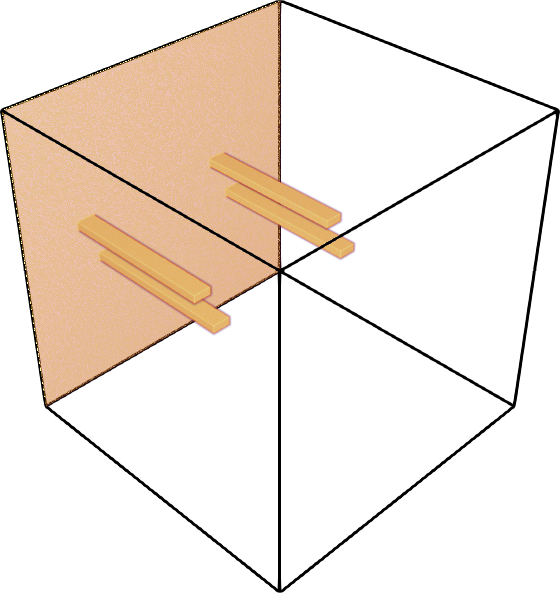}
            iii) stagger
        \end{minipage}
        \hspace{0.1\linewidth}
        \begin{minipage}{0.4\linewidth}
            \centering
            \includegraphics[width=\linewidth]{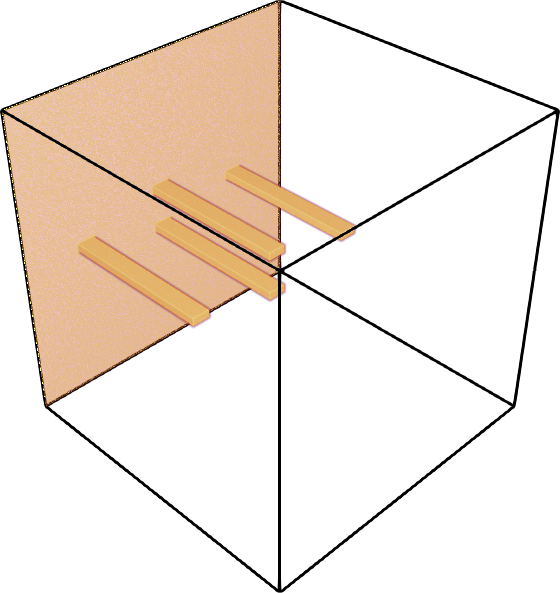}
            iv) star
        \end{minipage}\\
        \vspace{0.5em}
        \begin{minipage}{0.4\linewidth}
            \centering
            \includegraphics[width=\linewidth]{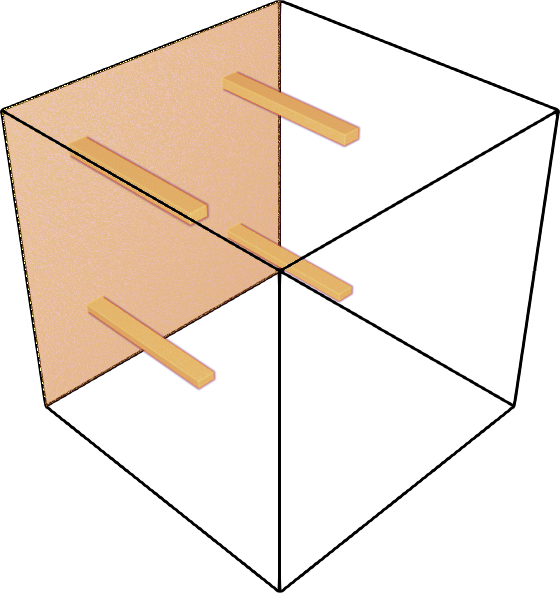}
            v) square
        \end{minipage}
        \hspace{0.1\linewidth}
        \vspace{0.5em}
        \begin{minipage}{0.4\linewidth}
            \centering
            \includegraphics[width=\linewidth]{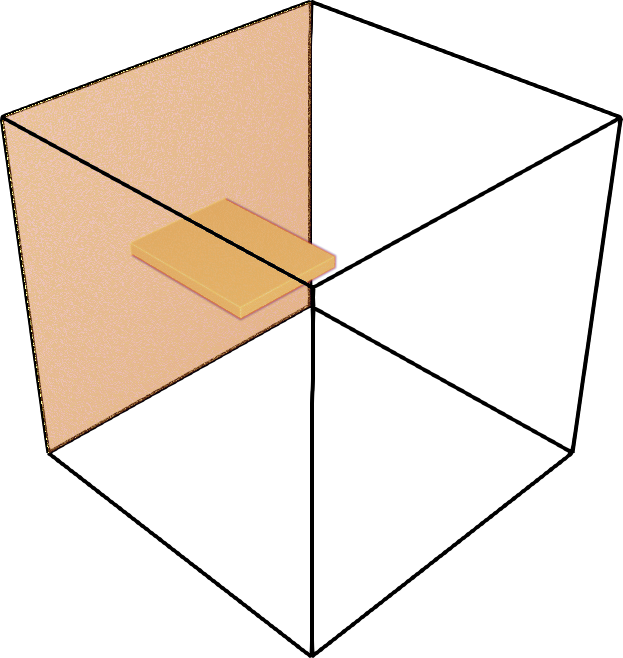}
            vi) midplate
        \end{minipage}\\
        \begin{minipage}{0.4\linewidth}
            \centering
            \includegraphics[width=\linewidth]{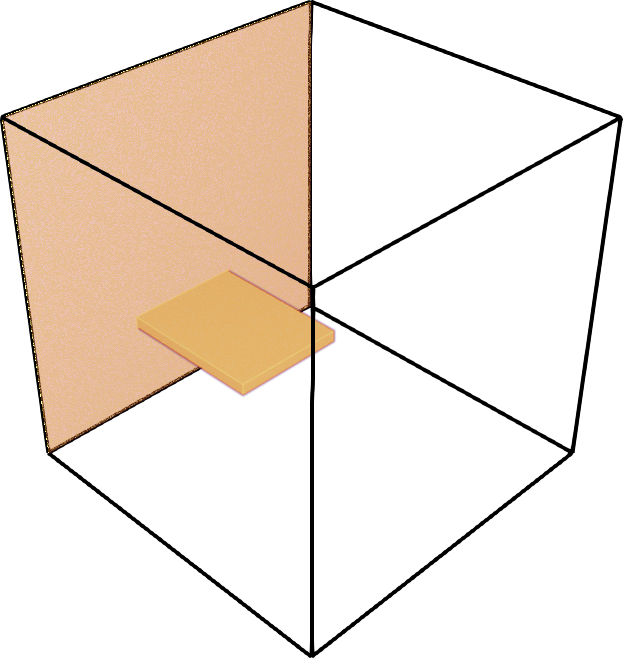}
            vii) lowplate
        \end{minipage}
        \hspace{0.1\linewidth}
        \begin{minipage}{0.4\linewidth}
            \centering
            \includegraphics[width=\linewidth]{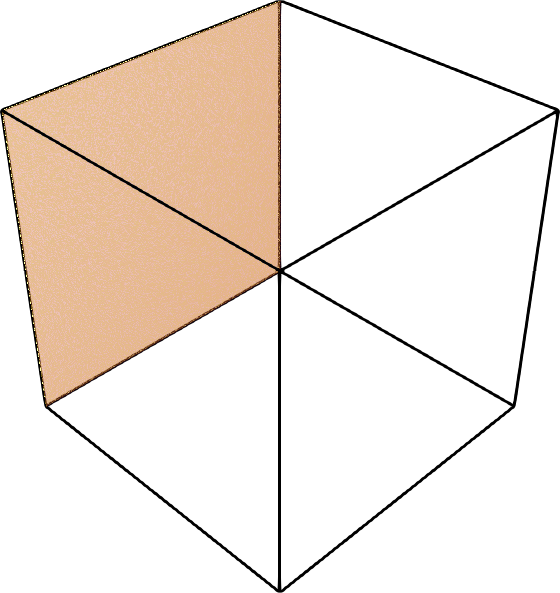}
            viii) finless
        \end{minipage}
    \end{subfigure}
    \caption{Panel (a): typical snapshots of temperature field in the melting dynamics of a PCM cell consisting of a cubic enclosure with side $L$ under the effect of gravity $\vec{g}=-g \hat{z}$. The solid substance is rendered as transparent. A side boundary of the enclosure consists of a hot wall at temperature $\Th=1$ featuring hot fins/plates protruding into the enclosure. Different dimensionless times $t^\star$ are reported (see text for details). Panel (b): different geometrical configurations used in the simulation. Panels (c)-(d): close up on the geometrical structure of the fins/plates used.}
    \label{fig:1}
\end{figure*}

Due to their large latent heat, Phase Change Materials (PCMs) prove exceptionally useful in storing heat in many fields, including cold energy storage~\cite{chavan-comprehensive-2022,altuntas-comprehensive-2025}, building insulation~\cite{wang-critical-2022}, thermal management in electronics~\cite{xie-investigation-2025} and hydride thermal management for hydrogen adsorption/release~\cite{facci-optimized-2021,maggini-increasing-}, to cite but few examples. The dynamics of phase change in a PCM system involves a complex coupling between buoyancy forces, melting and thermal convection~\cite{du-physics-2024,proia-melting-2024,oskouei-closecontact-2024,yan-natural-2023,wang-equilibrium-2021}: a solid substance melts due to the heating from a source boundary; the liquid, then, develops buoyancy-driven thermal convection, eventually leading to the complete liquefaction of the solid phase (see~\cref{fig:1}). \\
Since PCM heat transfer properties are usually suboptimal, many research efforts have been devoted to relieve this flaw. Among the many design solutions in the literature~\cite{togun-critical-2024,michalrogowski-recent-2023}, the insertion of fins delivers a simple, both conceptually and manufacturing-wise, yet effective way of transferring heat to a PCM. Paramount to fin effectiveness is their strategic placement, to promote diffusion in the early stages of melting, as well as convection in the late dynamics~\cite{wu-phase-2021,xie-investigation-2025,jaberi-optimizing-2025,bhowmik-embedding-2025,proia-heat-2025}. Recent examples in the literature include~\cite{xie-investigation-2025}, in which the authors study and optimize branching fin structures to boost the performance of a battery pack thermal management system based on PCMs;  in~\cite{jaberi-optimizing-2025}, a parametric study was conducted focusing on fin orientation and porosity, while in~\cite{bhowmik-embedding-2025},
the effect of fin shape was investigated. In \cite{dong-novel-2024}, honeycomb fins were investigated: the trade-off between the increase in system mass and volume associated with the fins is discussed, as well as the resulting improvement in hydrogen absorption/desorption rates; in \cite{shrivastav2024design}, a metal hydride–PCM reactor is optimized by introducing a ``fin factor'' index to balance heat transfer surface enhancement against geometric penalties, with the aim of improving thermally limited charging and discharging times. Further applications and optimizations can be found in~\cite{zhu-heat-2022,zhao-review-2022,amati2023enhancing}.\\
Even without buoyancy, melting is a complex simulation task, due to the presence of the moving boundary between the solid and liquid phases; including gravitational effects adds the non-linearity of convection to the evolution of the interphase boundary.
Thus, an accurate characterization of the temporal evolution of the melting front requires detailed local information. While such insight can be (partially) obtained through carefully designed experiments~\cite{huber-lattice-2008,wang-equilibrium-2021}, numerical simulations grant deeper insight, allowing higher spatial and temporal resolution for the evaluation of PCM performance~\cite{yan-natural-2023,wang-ice-2021,favier-rayleigh-2019,shyy-computational-2007}.
On the top of this quite intricate scenario, the addition of fins modifies the boundary shape adding another layer of complexity. \\
In this letter, we conduct numerical simulations to provide a quantitative insight on the impact of different geometrical configurations of fins in the enclosure, extending to three dimensions previous analyses conducted with the same numerical approach~\cite{proia-heat-2025}. In particular, we address key issues on the importance of the number of adopted fins, as well as their spatial layout. Our systematic analysis hinges on the study of equivalent configurations, i.e. layouts with the same heating areas but with different numbers of heating elements and different geometrical designs. The parametric tests are based on non-dimensional numbers characterizing the flow, providing a useful, technical operational tool. Thus, in order to dissect the effective role of fluid properties and heating element layouts, we also test the same configurations in different convection regimes (i.e. higher or lower relative convection strength), regulated by different values of the Rayleigh Number $\ra$. Later, we provide an analysis in the conductive regime for different values of the Stefan Number (i.e. the ratio of sensible heat to latent heat) $\ste$.
\section{Hydrodynamic description of thermal convection with melting}\label{sec:methods}
The PCM cell is modeled as a cubic enclosure of side length $L$ initially filled with a solid material at temperature $\Tc$. The solid material is heated from a hot source on the left wall at temperature $\Th$; the a side wall also presents different heating elements (multiple fins or a plate) protruding into the bulk of the enclosure (see~\cref{fig:1}). Heat transfer is triggered by the hot source, generating the melting of the solid material and triggering the convective dynamics of the molten phase~\cite{huber-lattice-2008,favier-rayleigh-2019,rabbanipouresfahani-basal-2018}. This complex scenario is studied at the hydrodynamical level via the continuum fields of density $\rho=\rho(\vec{x},t)$, velocity $\vec{u}=\vec{u}(\vec{x},t)$ and temperature $T=T(\vec{x},t)$; melting is locally monitored via the liquid fraction $\phi=\phi(\vec{x},t)$ that assumes values between $0$ (solid) and $1$ (liquid). These fields obey the Navier-Stokes equations with buoyancy forces plus an advection-diffusion equation for the temperature featuring a melting term~\cite{proia-melting-2024,huber-lattice-2008,guo-coupled-2002,chandrasekhar-hydrodynamic-2013}:
\begin{gather}
    \partial_t\rho+\vec\nabla \cdot \left(\rho \vec{u}\right)=0\label{eq:ns-mass}\\
    (\partial_t+\vec{u}\cdot\vec\nabla)\vec{u}=\nu\nabla^2\vec{u}-\frac{\vec\nabla p}{\rho}- \alpha(T-T_0)\vec{g}\label{eq:ns-momentum}\\
    \partial_t T +\vec{\nabla}\cdot(T \vec{u})= \kappa\nabla^2T - \frac{L_f}{C_p}\partial_t\phi   \label{eq:temperature-equation}
\end{gather}
where $p=p(\vec{x},t)$ represents the fluid pressure, $\nu$ the fluid kinematic viscosity, $\kappa$ the thermal diffusivity, $L_f$ the latent heat of fusion and $C_p$ the specific heat. The buoyancy term in the r.h.s. of Eq.~\eqref{eq:ns-momentum} is written in the Boussinesq approximation~\cite{chandrasekhar-hydrodynamic-2013,guo-coupled-2002,huber-lattice-2008}, being proportional to the thermal expansion coefficient $\alpha$, the temperature difference w.r.t. a reference temperature $T_0$, and the gravity acceleration $\vec{g}$ pointing in the negative $z$ direction (see Fig.~\ref{fig:1}). The melting term in the r.h.s. of Eq.~\eqref{eq:temperature-equation} models heat transfer due to the melting of the substance. The liquid fraction $\phi$ depends on the local enthalpy $h=h(\vec{x},t)$ via a a smoothed step function~\cite{shyy-computational-2007,huber-lattice-2008}:
\begin{equation}
    \label{eq:phi-of-h}
    \phi(\vec{x},t)=
    \begin{cases}
        0                                & h<h_s           \\
        \frac{h(\vec{x},t)-h_s}{h_l-h_s} & h_s\le h\le h_l \\
        1                                & h>h_l
    \end{cases}
\end{equation}
where $h_s=C_p\Tc$ is the solid phase enthalpy and $h_l=h_s+L_f$ the liquid phase enthalpy. The local enthalpy, in turn, depends on both the temperature and the liquid fraction, $h(\vec{x},t)=C_pT(\vec{x},t)+L_f \phi(\vec{x},t)$. The bulk equations~\eqref{eq:ns-mass}-\eqref{eq:temperature-equation} are supplemented with boundary conditions for both $\vec{u}$ and $T$. From the thermal point of view the fins and the wall from which the fins protrude are sources with a prescribed fixed temperature $\Th$; all the other boundaries of the cubic enclosure are set as insulating (i.e. \textit{adiabatic}) walls. All solid boundaries are treated as no-slip boundaries for the velocity field.\\ The numerical simulations of~\eqref{eq:ns-mass}-\eqref{eq:temperature-equation} are performed by means of the Lattice Boltzmann models (LBM), a numerical tool that has proven to be reliable and accurate for the study of complex flow dynamics~\cite{succi-lattice-2001,kruger-lattice-2017,falcucci-extreme-2021,falcucci-adapting-2024,falcucci-hydrodinamic-2026} . More specifically, we developed a LB-based solver capable of simulating the convective flows coupled to melting dynamics in presence of complex boundary conditions~\cite{huber-lattice-2008,facci-optimized-2021,wang-ice-2021}. Here, the code  is a 3D extension of the code developed and validated in earlier studies by the authors, focused on the study of melting dynamics in 2D PCM cells~\cite{proia-melting-2024,proia-heat-2025}.
\section{Geometry and Setup}\label{sec:setup}
Different geometrical configurations of the PCM cell are analyzed.
The layouts are systematically selected in order to isolate the role of geometry, while keeping constant the total heated area and source temperature $\Th$. More specifically, the present analysis is aimed at providing a consistent three-dimensional framework to systematically assess widely used designs in the literature~\cite{sharifi-enhancement-2011,oskouei-closecontact-2024,zhu-heat-2022,chunrongzhao-fin-2022,huo-lattice-2017}.\\
The very first group of configurations stems from the simple idea of repeating a fin a number of times: four was chosen as it is not too high nor too low. A variety of diverse geometrical configurations is then obtained via translation operations, namely: i) \textit{Line}, with the 4 fins horizontally aligned at the same height; ii) \textit{Rect}, with two fins at the same height, the other two higher at the same horizontal position; iii) \textit{Stagger}, similar to rect, but with a slight horizontal offset for the higher fins; iv) \textit{Star}, with two fins at the same height, the other two at the same horizontal position, halfway between the other two, one higher and one lower. To better understand the role of the relative closeness of the fins, we also devised another geometrical arrangement: v) \textit{square}, a configuration in which the fins are at a larger, equal distance between themselves. The above configurations are shown in~\cref{fig:1}b.
The length, height and width (see~\cref{fig:1}c) for a single fin are $(l,h,w)=(63,4,8)/L$.\\
In order to address the effect of the number of fins, we also designed configurations with a single plate. The plate is designed with a total heating surface equal to that of the fins: given the heating surface of the fin, $S_\text{\tiny H,FIN}=2(lw+lh+wh)-wh$, the heating surface of the plate is taken as
$S_\text{\tiny H,PLATE}=4 \cdot S_\text{\tiny H,FIN}$. Note that in the computation of the heating surface, we have excluded the contact surface between the fin and the wall. The dimensions of the plate (see~\cref{fig:1}d) were then chosen as $(a,b,c)=(56,5,48)/L$.
Moreover, the inclusion of the plate was studied in two configurations, namely: vi) \textit{Midplate}, with a single, slab-like ledge positioned at about half the height of the enclosure; vii) \textit{Lowplate}: similar to \textit{Midplate}, but at a lower vertical position, in order to test the importance of the latter~\cite{proia-heat-2025} in 3D environments. Finally, to systematically assess the impact of the presence of the fins/plate, compared to a finless wall we have considered two further layouts: vii) \textit{Finless}, a simple cubic enclosure, with no fins, and a bare, side wall of area $L^2$ as the source at temperature $\Th$; viii) \textit{FinlessHotter}, similar to \textit{Finless}, but with a source temperature $\Thfl$. $\Thfl$ is chosen in such a way that its integral over the side boundary of area $L^2$ equals the integral of the source temperature $\Th$ in a configuration with fins, i.e. $\Thfl \cdot L^2= \Th \cdot(L^2+4 \cdot S_\text{\tiny H,FIN})$.\\
The performance of the different geometrical configurations is analyzed by changing also relevant non-dimensional numbers governing the phenomena under investigation.
To measure the intensity of buoyancy forces w.r.t. diffusive/dissipative effects, we use as an input parameter the Rayleigh Number $\ra$:
\begin{equation}\label{eq:rayleigh-number}
    \ra=\frac{g\alpha L^3}{\nu\kappa}(\Th-\Tc).
\end{equation}
The Stefan Number
\begin{equation}\label{eq:stefan-number}
    \ste=\frac{C_p}{L_f}(\Th-\Tc)
\end{equation}
accounts for the ratio of sensible to latent heat. The Prandtl Number $\pr=\nu/\kappa=1$ is kept fixed. To monitor the melting process within the cell we take the space integral of the liquid fraction normalized to the PCM volume $V_{\text{PCM}}=L^3-4\cdot whl$
\begin{equation}\label{eq:phi-star}
    \langle \phi \rangle (t)=\frac{\sum_{\vec{x}}\phi(\vec{x},t)}{V_{\text{PCM}}}.
\end{equation}
Finally, for the sake of simplicity, the results are reported using the dimensionless time $t^\star=t/t_{\text max}$, with $t_{\text max}=2\cdot10^4$ for all the simulations. The other parameters are fixed, in lattice units, at $g=9.8067,\,\nu=0.1,\,\kappa=0.1,\,L=128,\,T_H=1,\,T_C=0,\,T_0=0.5,\,\Thfl=1.3890,\,L_f=10^3$.
\section{Results and discussions}\label{sec:results}
\begin{figure}[t!]
    \centering
    \includegraphics[width=\linewidth]{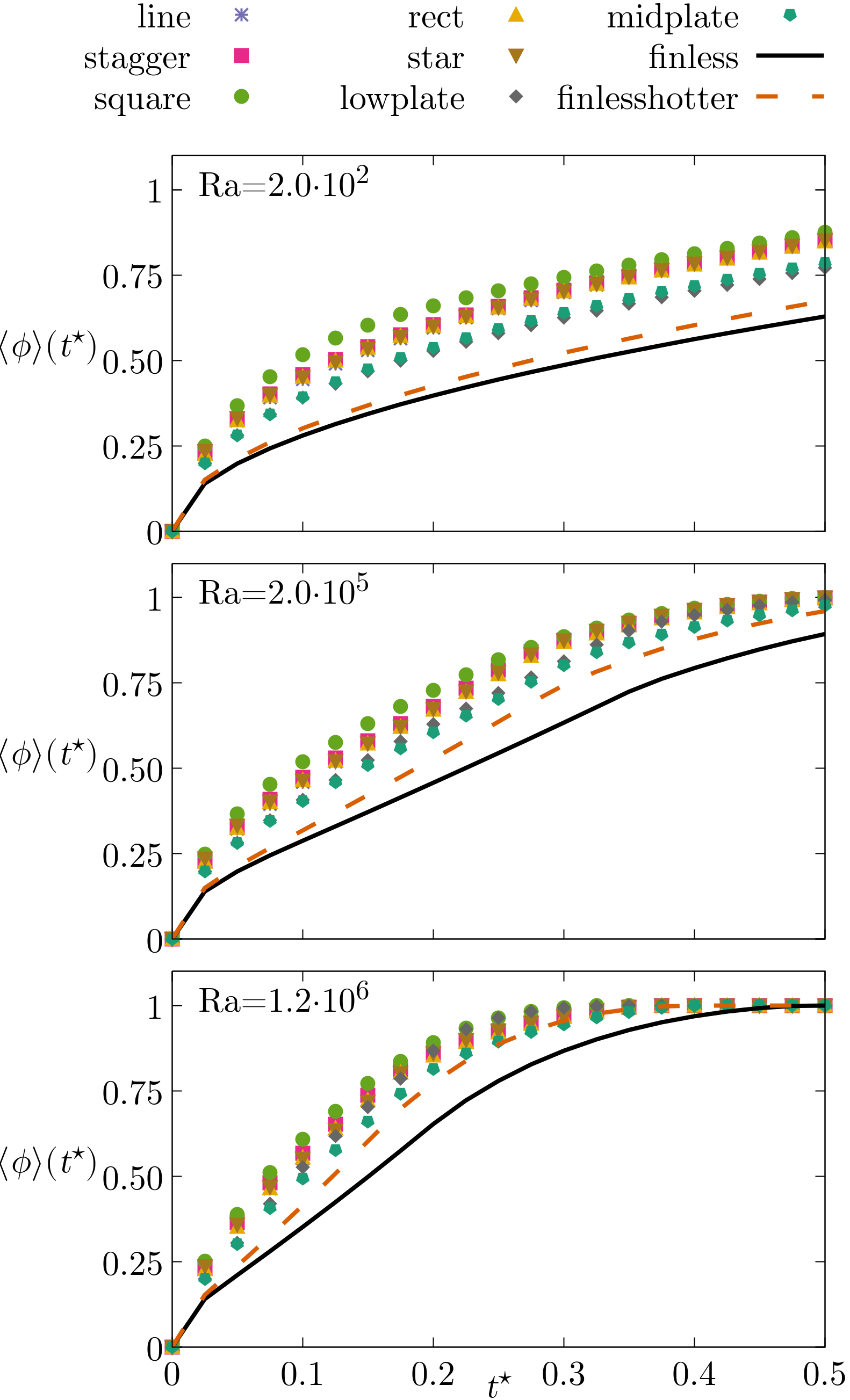}
    \caption{We report the normalized liquid fraction (cfr.~\cref{eq:phi-star}) as a function of the dimensionless time $t^\star$ for different Rayleigh numbers (cfr.~\cref{eq:rayleigh-number}) for a fixed Stefan number $\ste=10$. The different geometrical configurations are illustrated in~\cref{fig:1} with the exception of finlesshotter whose details are given in the text.}
    \label{fig:ra-comparison}
\end{figure}
\begin{figure*}[t!]
    \centering
    \captionsetup[subfigure]{singlelinecheck=off,justification=raggedright}
    \begin{subfigure}[t]{0.48\linewidth}
        \caption{}
        \vspace{1em}
        \centering
        \includegraphics[width=\linewidth]{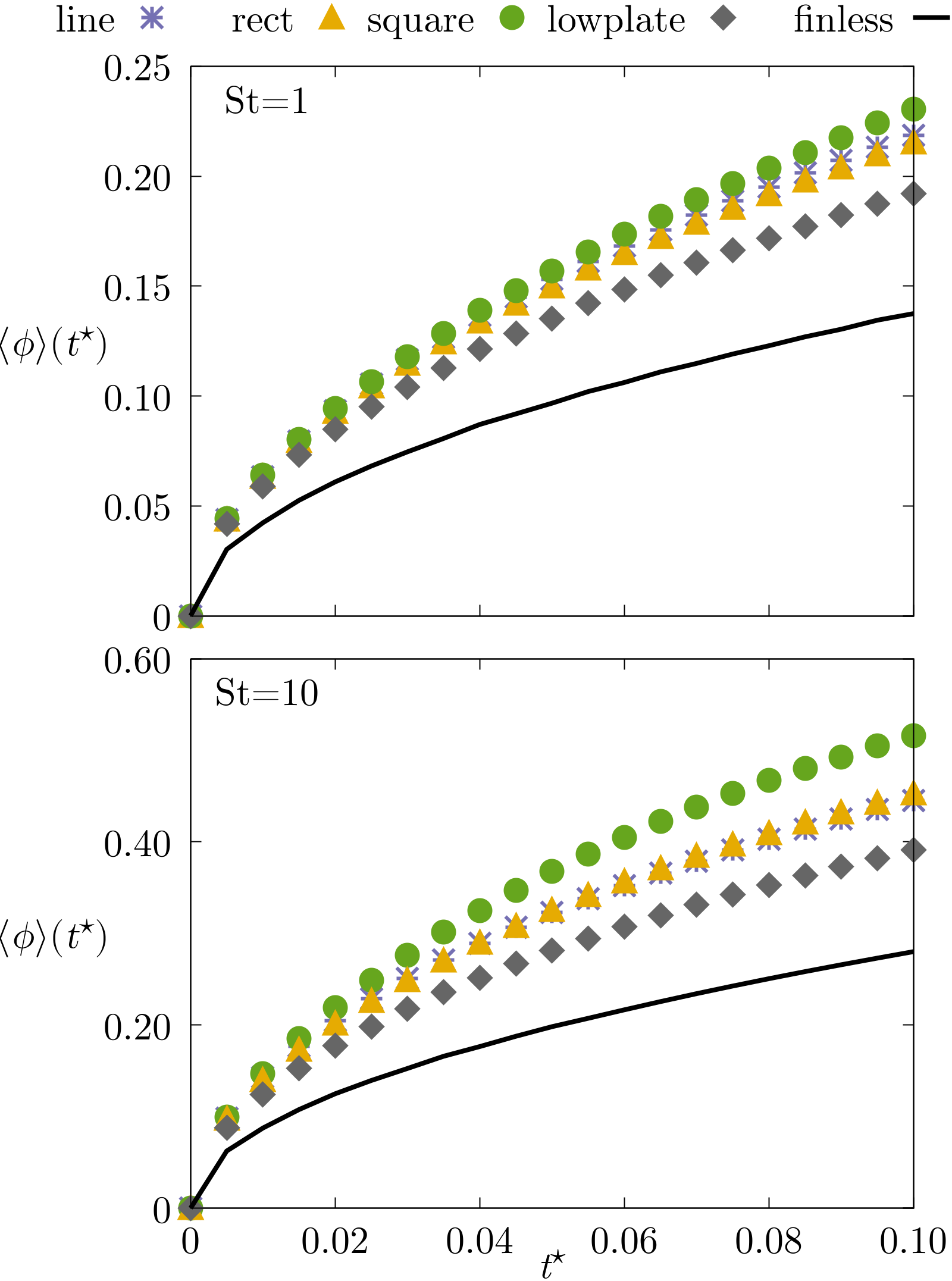}
    \end{subfigure}
    \hfill
    \begin{subfigure}[t]{0.48\linewidth}
        \caption{}
        \centering
        \begin{minipage}{0.45\linewidth}
            $$t^\star=1.5\cdot10^{-2}$$
            \begin{minipage}{\linewidth}
                \includegraphics[width=0.85\linewidth]{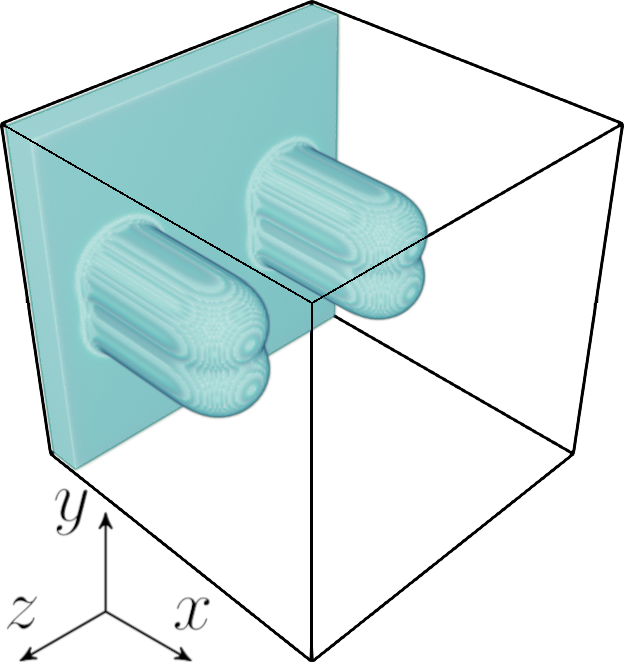}
                \hspace{-4em}
                \includegraphics[width=0.40\linewidth]{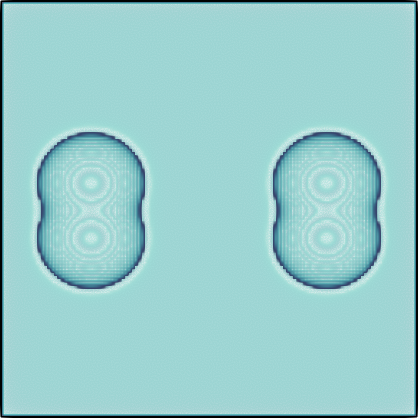}
            \end{minipage}\\
            \vspace{1em}
            \begin{minipage}{\linewidth}
                \includegraphics[width=0.85\linewidth]{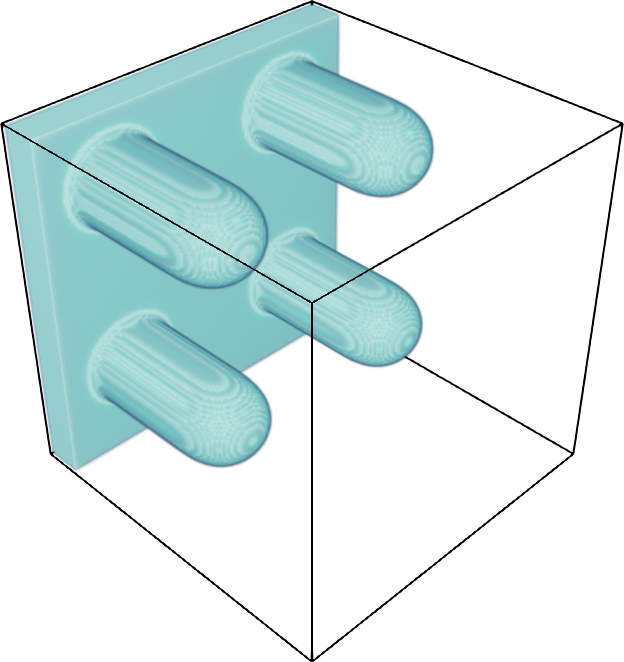}
                \hspace{-4em}
                \includegraphics[width=0.40\linewidth]{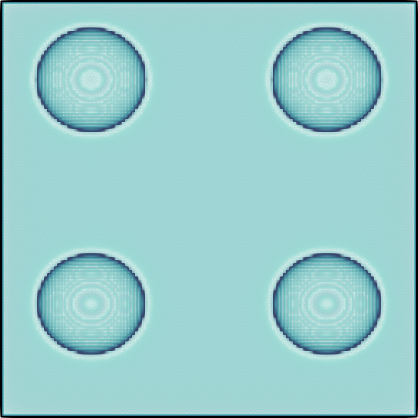}
            \end{minipage}\\
            \vspace{1em}
            \begin{minipage}{\linewidth}
                \includegraphics[width=0.85\linewidth]{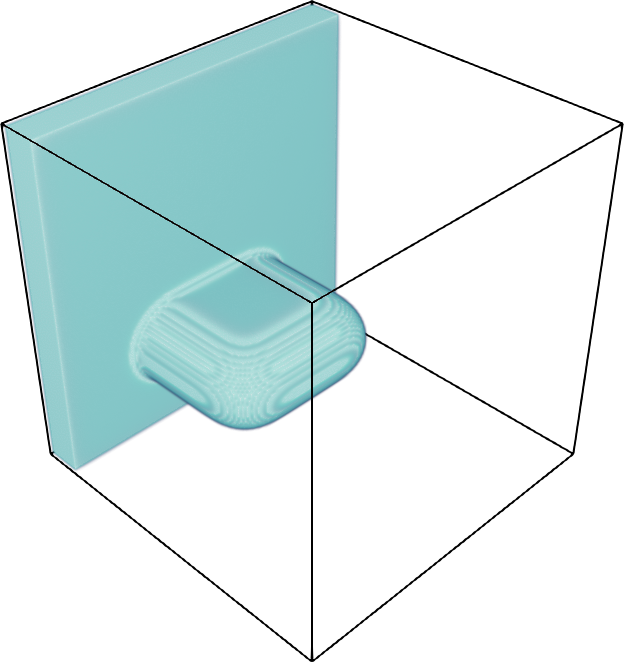}
                \hspace{-4em}
                \includegraphics[width=0.40\linewidth]{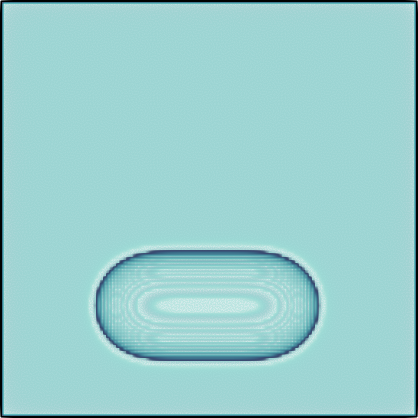}
            \end{minipage}
        \end{minipage}
        \hfill
        \begin{minipage}{0.45\linewidth}
            $$t^\star=1.0\cdot10^{-1}$$
            \begin{minipage}{\linewidth}
                \includegraphics[width=0.85\linewidth]{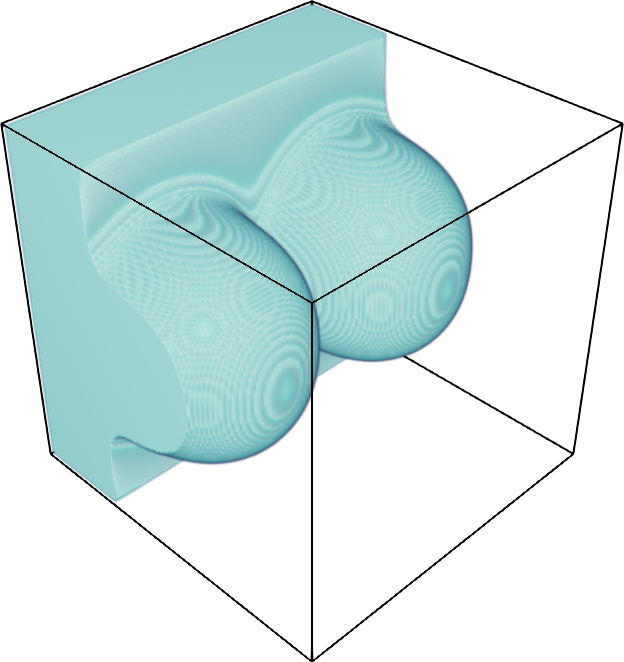}
                \hspace{-4em}
                \includegraphics[width=0.40\linewidth]{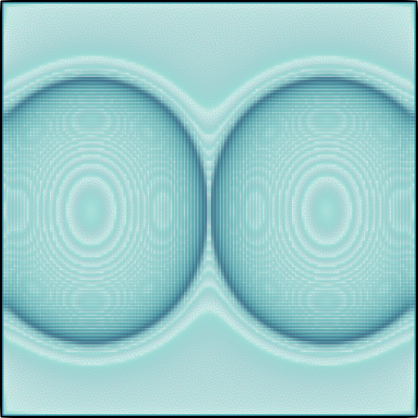}
            \end{minipage}\\
            \vspace{1em}
            \begin{minipage}{\linewidth}
                \includegraphics[width=0.85\linewidth]{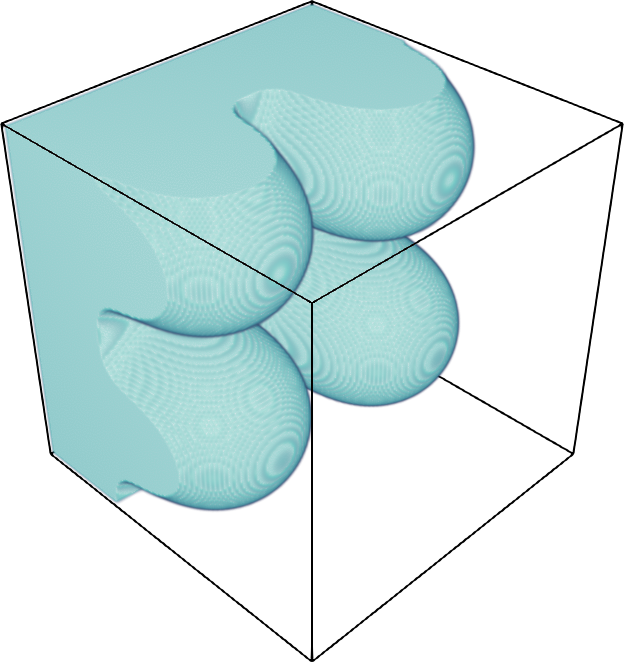}
                \hspace{-4em}
                \includegraphics[width=0.40\linewidth]{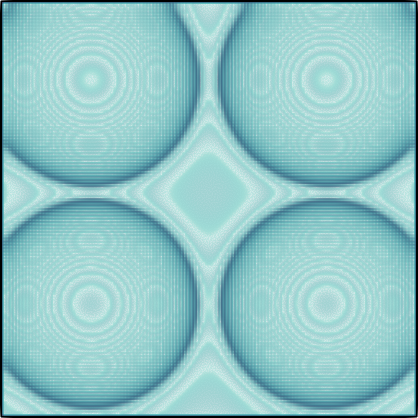}
            \end{minipage}\\
            \vspace{1em}
            \begin{minipage}{\linewidth}
                \includegraphics[width=0.85\linewidth]{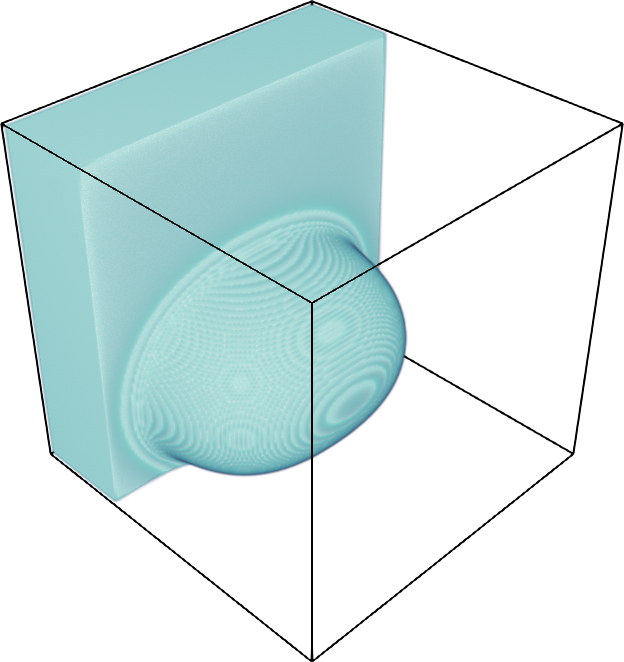}
                \hspace{-4em}
                \includegraphics[width=0.40\linewidth]{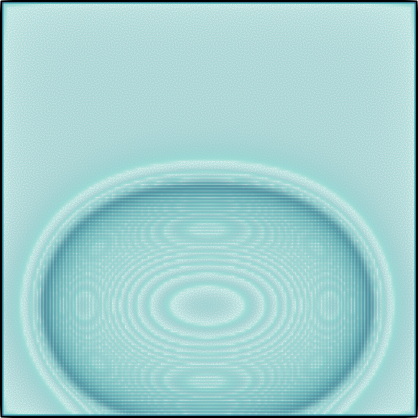}
            \end{minipage}
        \end{minipage}\\
        \vspace{1em}
        \begin{minipage}{\linewidth}
            \centering
            \includegraphics[width=\linewidth]{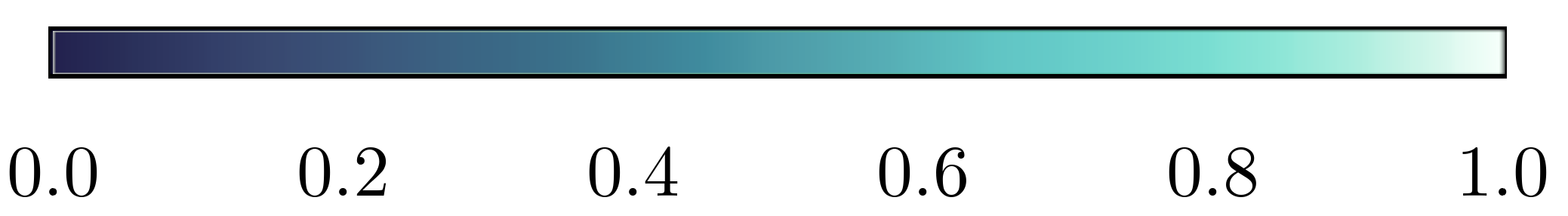}\\
            \vspace{-0.5em}$\phi$
        \end{minipage}
    \end{subfigure}
    \caption{Analysis of the early-stage melting dynamics in conductive case, corresponding to a zero Rayleigh number (cfr.~\cref{eq:rayleigh-number}). Panel (a): we report the normalized liquid fraction (cfr.~\cref{eq:phi-star}) as a function of the dimensionless time $t^\star$ for different geometrical configurations and different Stefan numbers (cfr.~\cref{eq:stefan-number}). The geometrical configurations used are illustrated in~\cref{fig:1}. Panel (b): 3D snapshots of the liquid fraction $\phi(\vec{x})$ at different times for the \textit{rect} and \textit{square} configurations, $\ste=10$. Each plot includes a front facing perspective on the bottom right.}
    \label{fig:fine-comparison}
\end{figure*}
First, we focus on the characterization of the time evolution of the normalized liquid fraction, according to~\cref{eq:phi-star}, as a function of $\ra$, while keeping $\ste = 10$, i.e. $C_p=10^4$. Figure~\ref{fig:ra-comparison} reports the trends of $\langle\phi\rangle(t^\star)$ for the different geometrical configurations illustrated in~\cref{fig:1}b, for $\ra = 2.0 \cdot 10^2, \ 2.0 \cdot 10^5$, $1.2 \cdot 10^6$, corresponding to $\alpha=10^{-7},10^{-4},6\cdot10^{-4}$. Increasing $\ra$ results in better performance for all the configurations in terms of reduced time to achieve the complete melting ($\langle \phi \rangle=1$) of the solid PCM. This observation is in line with the literature, as it is known that higher values of $\ra$ point to increased importance of convection, which is the main driver of heat transfer in melting systems~\cite{yang-morphology-2023,jany-scaling-1988,du-sea-2023,wang-Numerical-2017}. Moreover, simply raising the source temperature proves less effective than inserting fins, as \textit{finlesshotter} is generally worse-performing than the finned configurations, proving the fundamental role of the geometry. Increasing $\ra$ can reduce the advantage of the finned configurations with respect to other configurations, as the fins can impede convection by breaking up macroscopical flows; in fact, a \textit{proper} positioning of the fins in the lower parts of the enclosure can favor melting by promoting the onset and development of convection~\cite{proia-heat-2025}.

Regarding the performance of the plated configurations, we can see that {\it lowplate} performs generally better than {\it midplate}. We also notice that this is more evident at higher $\ra$, pointing to the same advantage of lower positioning that favors the development of larger convective structures~\cite{proia-heat-2025}

Results in Figure~\ref{fig:ra-comparison} exhibit a difference in performance between the plated and the finned configurations: we  argue that this is due to the different spatial distribution of the heating surface (which is always kept constant), that favors the performance of the finned configurations. We can also provide a simple but efficient argument to understand such a difference in the early stages of melting: one can assume that the substance melts at a characteristic distance $d$ from the heating surfaces. If we compute the difference between the substance melted by fins $V_{m,fin}$ and by the plate $V_{m,plate}$ we get\footnote{Choosing $t=0$ for the sake of simplicity}:
\begin{gather*}
    \Delta V_m=4V_{m,fin}-V_{m,plate}\\
    V_{m,fin}=(d+l)(2d+h)(2d+w)-lhw\\
    V_{m,plate}=(d+a)(2d+b)(2d+c)-abc
\end{gather*}
where $lhw$ and $abc$ are the volumes of the heating elements (we are excluding the wall-facing surface). Substituting $L$ and the chosen fin and plate dimensions:
$$
    \Delta V_m=\frac{3}{64}d^2 (256 d + 129),
$$
which is strictly positive. This computation neglects the melting from the side boundary which is reasonably assumed equal in both cases. The positivity of $\Delta V_m$ is an indication that the finned configuration uses the \textit{same} heating surface more efficiently, melting more substance in the same time. Of course, we must stress that this computation is reasonable only in the early stages of melting, where conduction dominates the dynamics. Nevertheless, the fact that the finned configurations produce a larger molten region in the early stages,  increases the effective liquid–solid interfacial area. This, in turn, promotes heat transfer and leads to a sustained advantage in the overall melting dynamics.\\
Regarding the relative differences between plated and finned configurations, we note that by increasing $\ra$, the plated configurations get closer to the finned ones, i.e. the latter configurations gain more (in terms of melted PCM) compared to the former; arguably, this is due to the presence of a larger surface $\perp\vec{g}$, since we have $ac>4lw$ for the chosen sizes.

Another important feature to remark in the analysis of Figure~\ref{fig:ra-comparison} is that configurations i)-iv), despite being quite geometrically different, deliver essentially the same performance. Configuration v) (square) with further apart fins, performs better than i)-iv). This suggests the presence of effects triggered by the closeness of the fins that may lead to an energetic overlap in molten zones, during the early stages, thus ``wasting'' thermal energy on already melted substance. In order to provide some quantitative evidence of this hypothesis, we focus on a purely conductive regime ($\ra=0$) to be able to stress the difference between v) and i)-iv). Moreover, since the melting speed in the conductive dynamics is regulated by the Stefan number $\ste$, we analyze data related to two different values of $\ste$, namely $\ste=1,10$, obtained by choosing $C_p=10^3,10^4$, aiming to prove the robustness of the trend. In~\cref{fig:fine-comparison}a we analyze the dynamics of the normalized liquid fraction in the conductive regime, for different geometrical configurations. A lower value of $\ste$ delivers slower melting, but the separation happens regardless, just delayed in time. Again, the plated configurations perform worse. In~\cref{fig:fine-comparison}b snapshots of $\phi(\vec{x})$ for $\ste=10$ are shown for \textit{rect} and \textit{square} at the time for which the latter visibly separates from the other configurations ($t^\star=1.5\cdot 10^{-2}$) and at the last time shown in~\cref{fig:fine-comparison}a ($t^\star=1.0\cdot 10^{-1}$). A comparison between the first and second rows highlights that an interference in the molten zones can already be seen for the \textit{rect} configuration at times for which it is absent for the \textit{square} configuration ($t^\star=1.5\cdot 10^{-2}$). This overlap is clearly responsible for the reduction of the total molten volume in the \textit{rect} configuration. In the third row, the \textit{lowplate} configuration is also shown, giving visual proof of the worse spatial efficiency of such configurations. Overall, the better performance of \textit{square} proves that the relative distance between fins matters: further apart fins gain an advantage that cascades in overall better melting.
\section{Conclusions}\label{sec:conclusions}
Phase Change Materials (PCMs) are playing an increasingly important role in energy-related applications, ranging from thermal energy storage and building technologies to thermal management in electronics and hydrogen systems~\cite{facci-optimized-2021,nivedhitha-advances-2024,maggini-numerical-2025,maggini-nondimensional-2024}. Improving PCM heat transfer capabilities without introducing active components is a key technological challenge, making passive optimization strategies particularly attractive for engineering applications. The insertion of fins inside a PCM is known to be a simple and effective way of enhancing the overall thermal performance~\cite{sharifi-enhancement-2011}. This letter systematically addressed the impact of fin layout and geometrical configurations on melting. First, two aspects already known from the literature~\cite{proia-heat-2025,sharifi-enhancement-2011} were confirmed: 1) fin insertion provides more efficient melting with lower relative temperature due to the onset of convection and 2) fins positioned in the lower parts of the chamber give rise to larger convective motions (and, thus, enhanced thermal transfer) than equivalent fins positioned in the upper part of the domain.
We further observed that distributing the same heating surface over multiple slender fins is more effective than concentrating it into a single plate-like element, as multiple sources melt more PCM in the early stages, leading to larger melting rates. From the analysis of the fin layouts, we found that the relative positioning of fins can lead to overlapping molten zones, due to excessive closeness: too-close configurations lead to early overlap of molten regions, resulting in inefficient use of thermal energy. In contrast, configurations with more distant fins are characterized by a smoother distribution of the heat, leading to efficient melting.\\
The findings in this letter pave the way for further investigations in order to assess whether they remain valid in a wider range of values of $\ra$. Furthermore, testing different sizes/shapes for the designed fins, as well as studying more in detail the role of interstitial spaces in relation to the fin height could reveal more interesting peculiarities on melting dynamics. Finally, exploring more complex fin geometries and resorting to hybrid technical solutions (metal foams, nanoparticle insertion and different chemical compositions among the others), to improve thermal transfer~\cite{gao-lattice-2017,bianco-finned-2021,zhao-review-2022,michalrogowski-recent-2023,togun-critical-2024} could help extend the present analysis towards broader design principles.
\acknowledgments
The authors acknowledge the support of the National Center for HPC, Big Data and Quantum Computing, Project CN\_00000013 - CUP E83C22003230001, Mission 4 Component 2 Investment 1.4, funded by the European Union - NextGenerationEU. Support from INFN/FIELDTURB project is also acknowledged.
The authors wish to acknowledge the support of Project PRIN 2022 F422R2-
CUP E53D23003210006, financed by the European Union-NextGenerationEU; the support of Project PRIN PNRR P202298P25 - CUPE53D23016990001,
financed by the European Union-NextGenerationEU; and the support
of Project ECS 0000024 Rome Technopole-CUP B83C22002820006, NRP
Mission 4 Component 2 Investment 1.5, funded by the European Union-
NextGenerationEU.
This work was partially supported by the Office of Naval Research of the United States of America, with Drs. J. Dibelka and S. Shroeder as program
managers (Grant N62909-24-1-2072).

\bibliographystyle{eplbib}
\bibliography{3d_multifin_bib}

\end{document}